\documentclass[runningheads]{llncs}
\usepackage[T1]{fontenc}
\usepackage{graphicx}
\usepackage{booktabs}
\usepackage[misc]{ifsym}
\newcommand{\corr}{(\Letter)}
\usepackage{mwe}

\begin{document}

\title{Modelling the longevity of complex living systems}

\author{Indr\.e \v{Z}liobait\.e \corr \orcidID{0000-0003-2427-5407}}

\authorrunning{I. \v{Z}liobait\.e}

\institute{University of Helsinki, Finland \email{indre.zliobaite@helsinki.fi}}

\maketitle              

\begin{abstract}
This extended abstract was presented at the Nectar Track of ECML PKDD 2024 in Vilnius, Lithuania. The content supplements a recently published paper "Laws of Macroevolutionary Expansion" in the Proceedings of the National Academy of Sciences (PNAS) \cite{Zliobaite24_pnas}.

\end{abstract}


Van Valen's Law of Constant Extinction \cite{VanValen73} postulates that in the same ecological contexts \cite{VanValen71}, the probability of extinction of a biological species over the next time step does not depend on how long the species already existed. In systems like this the durations of biological species commonly follow exponential decay \cite{Sole22}. Models like this, are also used to model breakdowns of unstable isotopes, durations of phone calls, or failures of mechanical systems. A key aspect in common across those examples is that entities are exposed to many independent trials, and a failure can happen in any of those trials. In macroevolution, this process is known under the Red Queen's hypothesis \cite{VanValen73,Zliobaite17}, which states that "the effective environment of any homogeneous group deteriorates at a stochastically constant rate" \cite{VanValen73}. "To a good approximation, each species <...> is a part of a zero-sum game against other species. Which adversary is most important for a species may vary from time to time, and for some or even most species, no one adversary may ever be paramount. Furthermore, no species can ever win, and new adversaries grinningly replace the losers" \cite{VanValen80}. A competitive system like this leads to species durations being distributed exponentially, where the distance between events follows a Poisson point process, in which events occur continuously and independently at a constant average rate. 

\begin{figure}[t]
\includegraphics[width=0.24\textwidth]{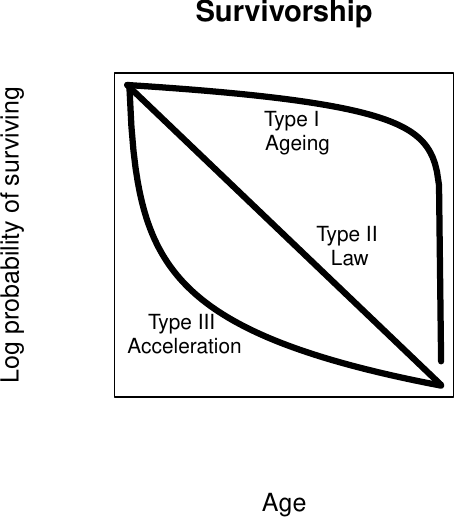}
\includegraphics[width=0.24\textwidth]{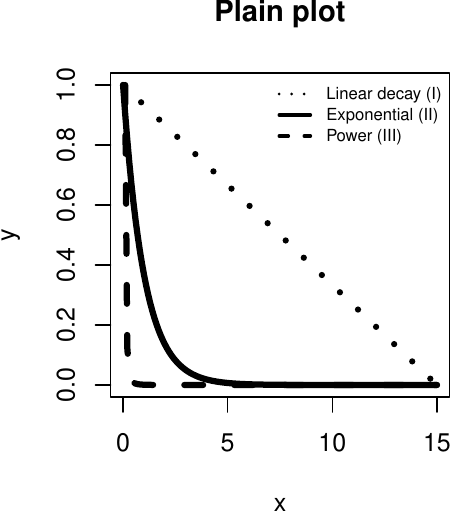}
\includegraphics[width=0.24\textwidth]{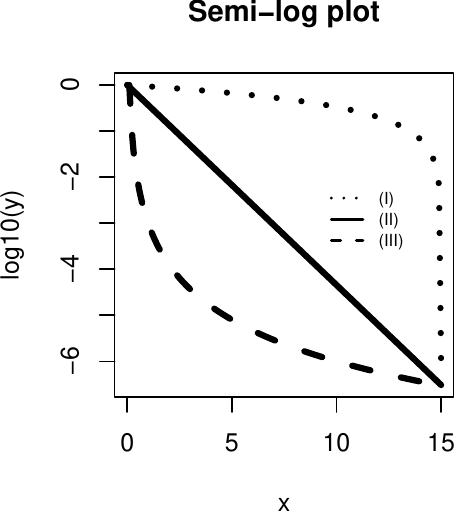}
\includegraphics[width=0.24\textwidth]{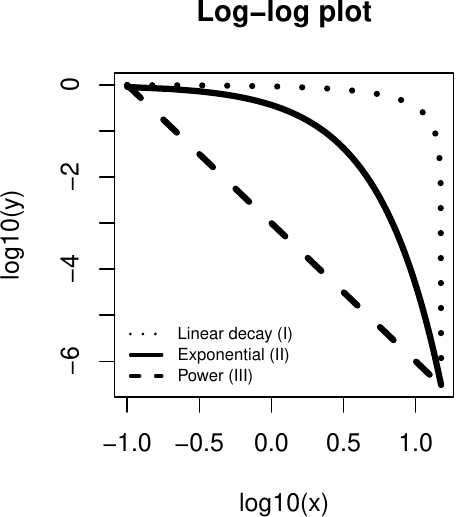}
\caption{Survivorship curves (a) and their mathematical models under three coordinate transformations: plain (b), semi-log (c) and log-log (d).} 
\label{Fig:1}
\end{figure}

Van Valen discovered the patterns of constant extinction when analysing the survivorship of taxonomic lineages in a way that has been common for demographic analysis of individuals \cite{Deevey47}, known as survivorship curves in ecology. When the proportion of individuals alive is plotted against their age on a semi-log plot, the shape of the curve can be used to reason about the probabilities of surviving from one age group to the next. The relationship can be of three types, as illustrated in Figure \ref{Fig:1}a. In a Type I relationship, the probability of dying in the next time step accelerates with an individual’s age. This follows from the slope of the survivorship curve being shallower at lower ages and steeper at higher ages. In a Type III relationship, the probability of dying decreases with age. This follows from the slope of the survivorship curve being steeper at lower ages. Type II relationship, which plots as a straight line on a semi-log plot, implies that the probability of dying in the next time step is the same for individuals of all ages. The mortality of individuals in a population rarely follows the Type II relationship \cite{Rauschert10}, yet Van Valen's  discovery showed that the mortality of taxonomic groups in the terrestrial and marine fossil record commonly does \cite{VanValen73}.

We propose a framework for modelling the three types of relationships mathematically, and testing for the best fitting relationship via different transformations of the coordinate system. 
Type I relationship corresponds to a linear decay model that takes the form $y = b - ax$ and gives a straight line in the plain coordinate system. 
Type II relationship corresponds to an exponential decay model that takes the form $y = ce^{-dx}$  and, a straight line in the semi-log plot. 
Type III relationship corresponds to a power law decay that takes the form $y = gx^{-h}$ and gives a straight line in the log-log plot. 
Here, $x$ is the input variable (taxon’s duration or range size), $y$ is the dependent variable (the proportion of taxa from the initial sample surviving or expanding at each step in time or space), $e$ is the Euler’s number, and $a$, $b$, $c$, $d$, $g$ and $h$ are parameters taking different values to fit the relationships to the data.

Figure \ref{Fig:1}b, c and d shows transformations of three base models in three coordinate systems: plain, semi-log and log-log. 

Type I survivorship curve is linear or closest to linear in the plain plot (Figure \ref{Fig:1}b). 
The number of taxa alive in this setting can be modelled as a linear regression $N_t = N_0  - ct$, $c$ is a parameter. 
The proportion of taxa vanishing at each time step is constant in relation to the initial number of species, then  
the probability of survival for a species to the next time step can be expressed as 
\begin{equation}
P_t^{\mathit{survival}} = 
N_t+\frac{1}{N_t} = 
\frac{N_0 - ct - c}{N_0 - ct} = 
1 - \frac{c}{N_0 - ct}.
\end{equation}
Here the probability of extinction depends on time $t$. 
If $t$ goes up, the second term will go up, thus the probability of survival will go down.  
Hence, if the data follows this relationship, extinction is not memoryless and implies aging.

Type II survivorship curve, which is linear in a semi-log plot (Figure \ref{Fig:1}c), can be mathematically described as an exponential decay. 
Exponential decay is typically used to model physical, chemical, or social processes or the decay of radioactive isotopes.
The amount of substance decreases at a constant rate in proportion to the amount remaining. 
This can be expressed via a differential equation as 
$\frac{dN}{dt} = -\lambda t.$
The solution to this equation is 
$N_t = N_0e^{-\lambda t},$
where $N_t$ is the amount of substance at time $t$; $N_0$ is the initial amount. 
Thus, when the number of taxa surviving over time follows this relationship, the probability of survival to the next time step can be expressed as
\begin{equation}
P_t^{\mathit{survival}} = 
N_t+\frac{1}{N_t} = 
\frac{N_0e^{-\lambda(t+1)}}{N_0e^{-\lambda t}} = e^{-\lambda}.
\end{equation}
We see that this probability does not depend on time $t$, and so the probability of extinction  $(1 - P_t^{\mathit{survival}})$ does not depend on $t$ either. The decay is memoryless.

Type III survivorship curve is linear or closest to linear in the log-log plot (Figure \ref{Fig:1}d). 
This corresponds to a power law, also known as the “rich get richer” or “80:20” rule.
The power law is often used to model phenomena that relate to behaviour, for instance, the distribution of wealth or the distribution of clicks or likes on the web. 
The power law is also commonly used to model the scaling of living organisms. A model form for power law survivorship is 
$N_t = bt^{-k}.$ In this case the probability of survival depends on time $t$, since 
\begin{equation}
P_t^{\mathit{survival}} = N_t+\frac{1}{N_t} = \frac{t+1}{t}^{- k}.
\end{equation}
We can see that if t goes up, the probability of survival will go up. This implies that the chances for survival improve with age.

This framework gives a simple way to test which type of a relationship observational data follows. Species durations can be analysed in three coordinate systems: plain, semi-log transformation, and log-log transformation to test which transformation gives the best linear fit. The best fit determines the closest type of a pattern, even in practice data can follow a combination of those relationships.

In this analysis we use Ordinary Least Squares regression to compare the coefficients of determination ($R^2$) of the three fits. 
While the absolute values of $R^2$s can turn out relatively high due to non-independence of the observation points, non-independence affects data similarly in all three transformations. 
Thus, relative comparisons of the statistics are meaningful. 

\begin{figure}[t]
\includegraphics[width=0.32\textwidth]{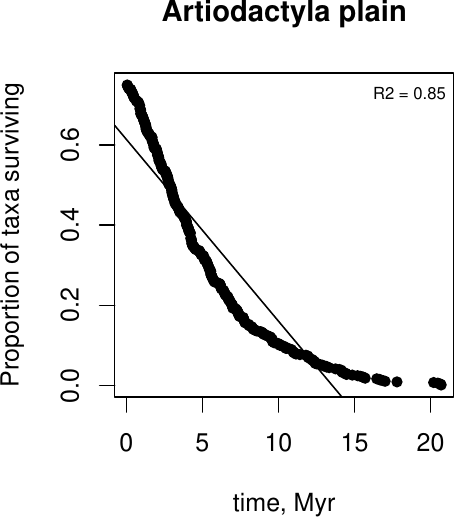}
\includegraphics[width=0.32\textwidth]{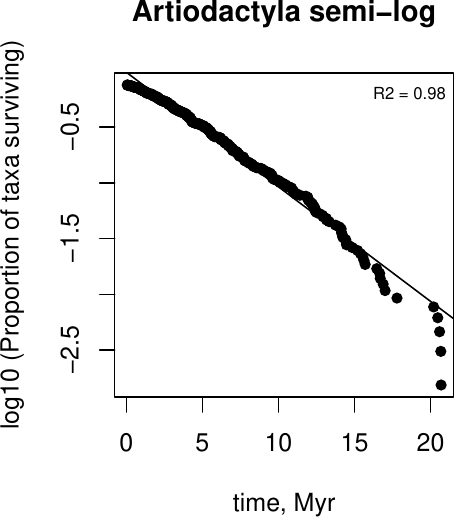}
\includegraphics[width=0.32\textwidth]{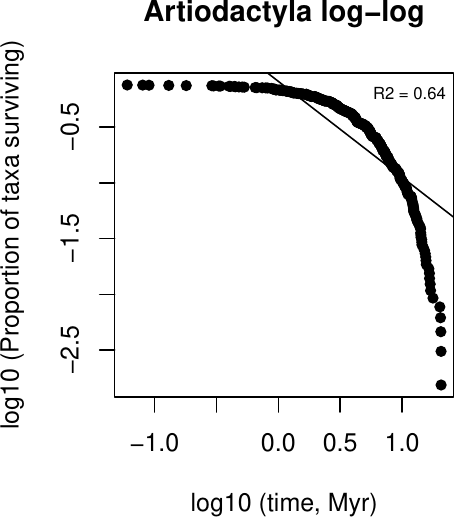}
\caption{Testing the type of survivorship for genera in the order Artiodactyla in the fossil record. The top right corner shows the coefficient of determination ($R^2$).} 
\label{Fig:2}
\end{figure}

Figure \ref{Fig:2} illustrates the test for survivorship of genera in the order Artiodactyla in the fossil record represented in the NOW database of fossil mammals~\cite{NOW}. 
This is the order of even-toed ungulates, such as antelopes, pigs, hippos, and more.  
Plots for other Mammalian orders are given in the online repository\footnote{\url{https://github.com/zliobaite/expansion}} accompanying the published paper \cite{Zliobaite24_pnas}. 
We see from Figure \ref{Fig:2} that the semi-log plot visually shows the best fit, as does the $R^2$ measure. 
From these results, we can conclude that the Law of Constant Extinction holds for Artiodactyls. 

The patterns we observe in the plots for other Mammalian orders mostly show Type II relationship as well, which is consistent with a memoryless macroevolutionary process postulated in the Red Queen's hypothesis. Memoryless extinction; however, does not mean that the ecology and the environment do not have any effect on which species will go extinct. 

Our proposed framework provides a simple way for testing the Law of Constant Extinction can also be used for analysing patterns of species expansion and decline in space \cite{Zliobaite24_pnas}. 

\begin{credits}
\subsubsection{\ackname} 
This is a contribution to the research projects: “Macroevolution of ecological relationships: does the key to the past wear out?” (Research Council of Finland, 354228) and “Comparing evolutionary processes in nature and society” (Kone Foundation). The contribution is from the Valio Armas Korvenkontio Unit of Dental Anatomy in Relation to Evolutionary Theory.

\subsubsection{\discintname}
The author has no competing interests in relation to this study. \end{credits}

%
%
\bibliographystyle{splncs04}
\bibliography{bib_nectar}

\end{document}